\documentstyle[aps,prl,multicol,epsf]{revtex}

\begin{document}

\title{ Self-heating in small mesa structures }

\author{ V.M.Krasnov,$^{1,2}$ A.Yurgens,$^{1,3}$
D.Winkler,$^{1,4}$ and P.Delsing$^{1}$}

\address{$^1$ Department of Microelectronics and Nanoscience,
 Chalmers University of Technology, S-41296 G\"oteborg, Sweden}

\address{$^2$ Institute of Solid State Physics, 142432 Chernogolovka,
Russia}

\address{$^3$ P.L.Kapitza Institute, 117334 Moscow, Russia}

\address{$^4$ IMEGO Institute, Aschebergsgatan 46, S41133, G\"oteborg,
Sweden}

\date{\today }
\maketitle

\begin{abstract}

We study analytically and numerically a problem of self-heating in
small mesa structures. Our results show that the self-heating is
proportional to a characteristic in-plane size of the mesa.
Experimental data for small high-$T_c$ superconductor Bi2212 mesas
are in qualitative agreement with our calculations. We estimate
the self-heating in Bi2212 mesas with different sizes and
demonstrate that the self-heating can effectively be obviated in
small mesa structures.

\end{abstract}

\begin{multicols}{2}

Progress in microelectronics results in a continuous decrease of
sizes of individual elements. Often, an active part of a
transistor is placed on top of a substrate, i.e., it has a "mesa"
geometry. Transport current through the mesa-structure may result
in large self-heating due to power dissipation in a small volume
and a poor thermal contact to the substrate. It is well known that
self-heating can considerably deteriorate the performance of
high-power transistors\cite{Liou,Bayrak}, especially at cryogenic
temperatures\cite{Cryo}. Therefore, details of thermal transport
in small mesa-structures is important for future development of
cryoelectronics.

High-$T_c$ superconducting (HTSC) Bi2212 mesa-structures are
promising candidates for
cryoelectronics\cite{Walenh,Latysh,Dong,Wang}. Due to the layered
structure of Bi2212, such mesas represent natural stacks of atomic
scale "intrinsic" Josephson junctions (IJJ's)\cite{Muller}.
Indeed, estimation of the stacking periodicity from Fraunhofer
modulation of Fiske steps\cite{Fiske} yields $s=15.5\AA$, in a
perfect agreement with the crystalline structure of Bi2212. The
atomic perfection of the naturally occurring tunnel junctions in
Bi2212 provides a reliable basis for cryo-devices giving a high
degree of homogeneity and reproducibility. IJJ's with their
record-large $I_c R_n$ values $\sim 5-10 mV$ are particularly
attractive for high frequency applications\cite{Walenh,Dong,Wang}
and have become a powerful tool for fundamental studies on HTSC
materials\cite{Schlen,Krasnov,Yurgens}. However, self-heating may
become an acute problem in
Bi2212-mesas\cite{SuzukiPRB,SuzukiIEEE,Gough} due to low thermal
conductivity.

In this letter we study self-heating in small mesa-structures both
analytically and numerically. An analytic solution for a
three-dimensional temperature distribution shows that self-heating
is proportional to the length of the mesa and that it can be
negligible in small Bi2212-mesas with a small number of IJJ's,
$N$.

The problem of self-heating in Bi2212 mesas has long been
anticipated. A pronounced back-bending (negative differential
resistance) of current-voltage characteristics (IVC's) is
typically observed at large bias. The back-bending increases (see
e.g., Fig.~10a in Ref.\cite{Schlen}) and the voltage separation
between quasiparticle branches in IVC's decreases with dissipated
power (e.g., see Fig.~3 in Ref.\cite{SuzukiPRB}). Such a behavior
is in qualitative agreement with progressive self-heating and/or
nonequilibrium quasiparticle injection in the mesas. A substantial
self-heating was found in short-pulse experiments\cite{SuzukiIEEE}
and direct mesa temperature measurements\cite{Gough}.

However, recently it was argued that self-heating is less
important in small mesas\cite{Latysh,Krasnov}. It was noted that
the back-bending of IVC's vanishes and the spacing between
quasiparticle branches remains constant (see eg., Fig.~3 in
Ref.\cite{Krasnov}) when the in-plane mesa size is decreased to a
few $\mu m$. Seemingly, this contradicts a simple estimation of
the overheating

\begin{equation}
\Delta T \simeq qt_c/\kappa_c, \label{Eq.1}
\end{equation}

which suggests that the self-heating should be independent on the
area of the mesa as the dissipated power density $q\sim j_c V$ is
constant for mesas with the same $j_c$. Here $t_c$ is the
thickness and $\kappa_c$ is the transverse thermal conductivity of
the Bi2212 single crystal, and $j_c$ is the $c$-axis critical
current density.

To show that the self-heating does decrease in small structures
let us consider a disk-like mesa with in-plane radius $a$, on top
of a bulk single crystal, see Fig.~1. We assume that the
stationary heat conduction is described by a heat diffusion
equation:

\begin{equation}
\nabla \cdot \kappa \nabla T = -\nabla \cdot q, \label{Eq.2}
\end{equation}

with anisotropic thermal conductivity, $\kappa_{ab}$ along the
$ab$-planes and $\kappa_c$ in the transverse direction ($c-$axis).
Assuming that the power is dissipated only inside the mesa,
boundary conditions become $\kappa_c
\partial T / \partial z = -q$ at the bottom of the
mesa, $z=0$, $|r| \leq a$, and $T=T_0$, at the bottom of the
crystal, $z=-t_c$.

In the crystal, Eq.~\ref{Eq.2} is reduced to the Laplace equation
by substitution $z^*=z \sqrt{\kappa_{ab} / \kappa_c}$. Then the
problem can be mapped to that of an electrical potential of a
charged conductor. The exact solution is known for a 3-axis
ellipsoid\cite{ECC}.  For a circular mesa on top of a
semi-infinite

\begin{figure}
\noindent
\begin{minipage}{0.48\textwidth}
\epsfxsize=0.9\hsize \centerline{ \epsfbox{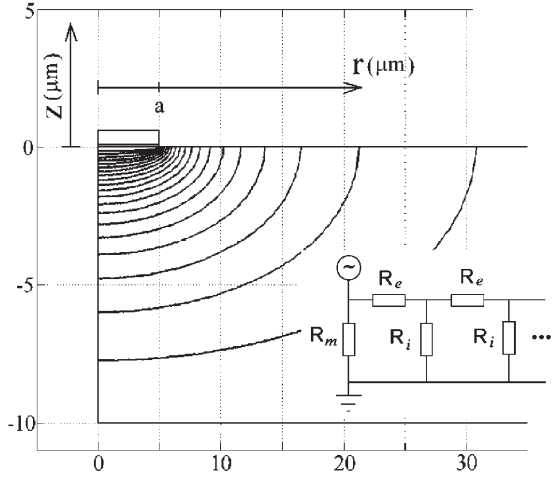} }
\vspace*{6pt} \caption{Results of numerical calculation with
self-consistent $\kappa(T)$. There are 19 isotherms in the
temperature range 10-41 K. }
\end{minipage}
\end{figure}

\noindent single crystal $t_c \rightarrow \infty $ the solution is

\begin{equation}
\Delta T_{\infty } = \frac{qa}{2 \sqrt{\kappa_{ab}\kappa_c}} atan
\left[ \frac{2a^2}{r^{*2}-a^2 + \sqrt{ \left( r^{*2}-a^2 \right)^2
+ 4a^2 z^{*2}}} \right]^{\frac 1 2}, \label{Eq.3}
\end{equation}

\noindent where $r^{ *2} = r^2+z^{*2}$ and $\Delta T = T - T_0$.

A solution for a finite $t_c$ can be obtained by an image method:

\begin{equation}
\Delta T(t_c) = \Delta T_{\infty}(r,z) - \Delta T_{\infty}(r, 2t_c
- z). \label{Eq.4}
\end{equation}

From Eqs.~(3,4), the self-heating of the mesa, $z=0$, $|r| \leq a$
is

\begin{equation}
\Delta T_m = \frac{q a}{2 \sqrt{\kappa_{ab}\kappa_c}} \left[ \frac
\pi 2 - atan \left( \frac {a }{2 t_c} \sqrt{\frac {\kappa_c} {
\kappa_{ab}}} \right) \right]. \label{Eq.5}
\end{equation}

Now it is clear that Eq.~\ref{Eq.1} is only justified for $a \gg 2
t_c \sqrt{\kappa_{ab}/\kappa_c}\approx 60\ \mu$m for a typical
$t_c \sim 10 \mu m$ and $\kappa_{ab}/\kappa_c \simeq 10$, i.e., in
none of practical cases.

For smaller mesas,

\begin{equation}
\Delta T_m \simeq \frac{\pi q a}{4 \sqrt{\kappa_{ab}\kappa_c}},
\label{Eq.6}
\end{equation}

\noindent which demonstrates that the self-heating decreases
linearly with a decrease of the in-plane size of the mesa.

To include the temperature dependence, $\kappa(T)$, we solved
self-consistently Eq.~\ref{Eq.2} using FEMLAB
package\cite{femlab}, taking empirical $k_{ab}(T)\approx
(1.06\times 10^{-3}T-7.57\times
10^{-6}T^2)$~Wcm$^{-1}$K$^{-1}$~\cite{Samoilov} and
$\kappa_{ab}/\kappa_c \simeq 10$~\cite{anisotropy}. Results of
numerical simulations for a disk-like mesa with the radius $a=5\
\mu$m on top of a single crystal 1~mm in diameter with $t_c =
10~\mu$m and $T_0=10$~K are shown in Fig.1 for a semi-space,
$r>0$. The numerical solution also revealed a linear decrease of
$\Delta T_m$ with $a$, but the self-heating was smaller due to an
increase of $\kappa$ with $T$. For example, in Fig.1 the maximum
temperature of the mesa is $\sim 41$~K for an empirical
$k_{ab}(T)$ and $\sim 67$~K for the case of
$k_{ab}=k_{ab}(T_0)=const$.

So far we assumed that the heat is sinking only into the pedestal
through the bottom surface of the mesa. In reality there is always
a metallic electrode on top of the mesa, which can help to cool
the mesa. Let's assume that the electrode with the thickness
$t_e$, width $w$, length $L_e$ and thermal conductivity
$\kappa_e$, is attached to the top surface of the mesa. The
electrode is separated from the Bi2212 by an insulating layer with
the thickness $t_i$ and the thermal conductivity $\kappa_i$. Even
though the thermal conductance of the electrode, $\kappa_e w
t_e/L_e$, is negligible due to the small cross-section and finite
length, it may yet provide effective $spreading$ of the heat
power. Effectiveness of such a heat-spreading metallic layer on
top of the mesa is known from the study of high-power
semiconducting transistors\cite{Bayrak}. To estimate the effective
thermal resistance $R_{th}=\Delta T /Q$, we once again use the
electric analog. The equivalent circuit is shown in inset of Fig.
1. Here $R_m \simeq (4a\kappa_{ab})^{-1}$ is the thermal
resistance of the pedestal, see Eq.(6), $R_e= \Delta x/(w t_e
\kappa_e)$ and $R_i= t_i/(w \Delta x \kappa_i)$ are thermal
resistances of an element $\Delta x$ of the heat-spreading layer
through the electrode and the insulating layer, respectively.
Because the thickness of the insulating layer is small, $t_i \sim
0.1 \mu m$, we used Eq.(1) for estimation of $R_i$. A
straightforward calculation yields

\begin{equation}
R_{th}(\Delta x \rightarrow 0) = \frac{R_m \sqrt{R_e R_i}}{R_m +
\sqrt{R_e R_i}}, \label{Eq.7}
\end{equation}

where $\sqrt{R_e R_i} = \left[t_i/(t_e w^2 \kappa_e \kappa_i)
\right]^{1/2}$ is the effective thermal resistance through the
electrode and insulator. Taking typical values of $\kappa_i = 0.1
W cm^{-1} K^{-1}$ (CaF$_2$ at low $T$), $\kappa_e = 1 W cm^{-1}
K^{-1}$ (Au), $t_e=t_i$ and $w=2a$ we obtain that $\sqrt{R_e R_i}$
can be an order of magnitude less than $R_m$. Even though this is
a rough estimation, it shows that the parallel heat channel
through the electrode can indeed be significant\cite{Note}.
However, whatever heat channel is dominating, the self-heating at
a given $q$ decreases with the in-plane size. Such a conclusion is
in agreement with earlier observations of better thermal
properties of small multi-emitter transistors as compared to
single-emitter transistors with the same total
area\cite{Liou,Bayrak}.

In Fig. 2, an experimental IVC at $T_0 = 4.2K$ is shown for a
$\sim 2 \times 2 \mu m^2$ Bi2212 mesa, containing a large amount
of IJJ's, $N \sim 200$. The Bi2212 is slightly overdoped with the
superconducting critical temperature, $T_c \simeq 90 K$. This data
is an example of extremely large power dissipation in a small
mesa. Indeed, because of large $N$, $q$ is more than $5 \cdot 10^4
\ W/cm^2$ at the largest bias. The self-heating is seen from the
back-bending of the IVC at large bias. On the other hand, multiple
quasiparticle

\begin{figure}
\noindent
\begin{minipage}{0.48\textwidth}
\epsfxsize=0.9\hsize \centerline{ \epsfbox{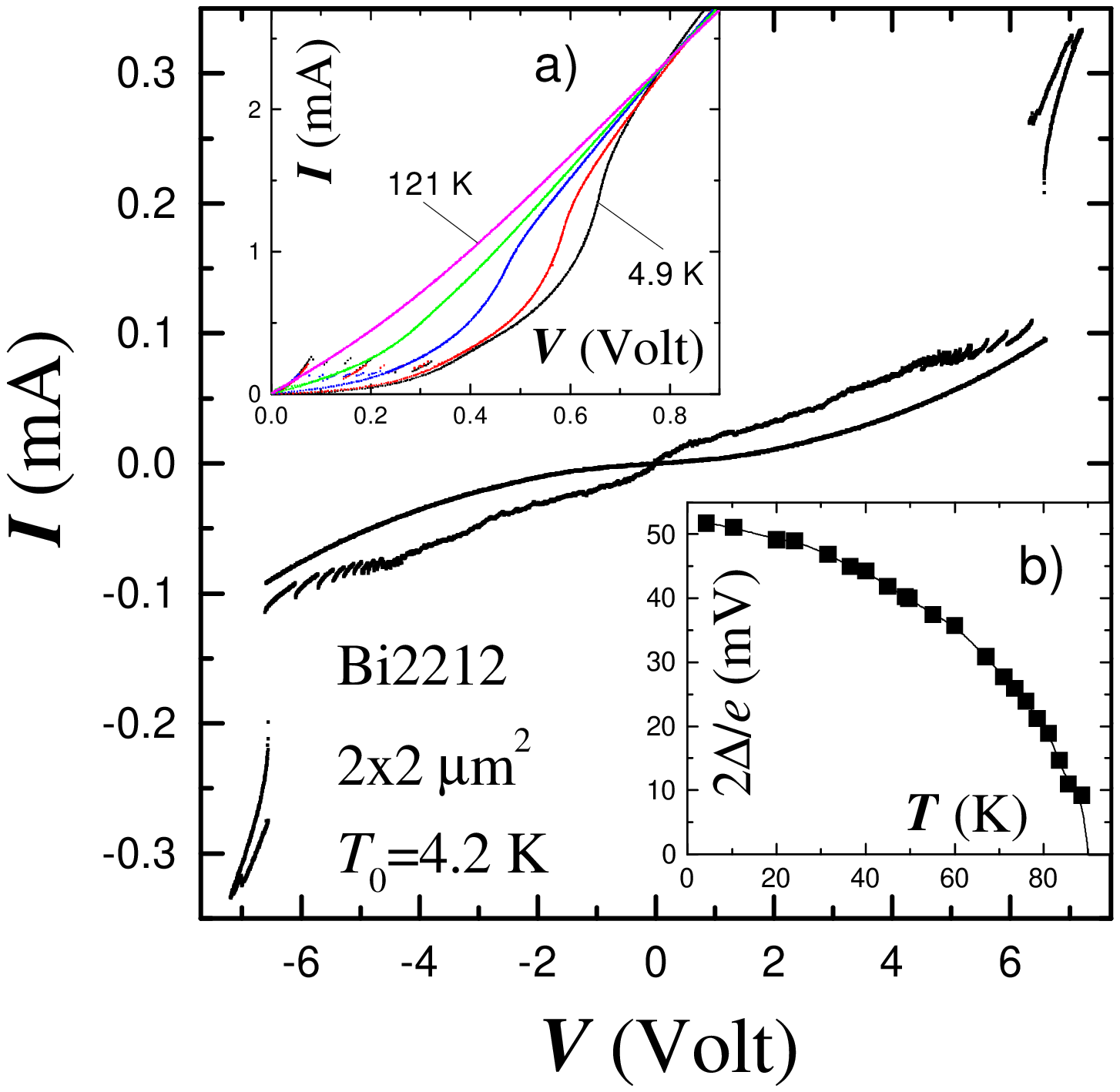} }
\vspace*{6pt} \caption{Experimental IVC's of $2 \times 2 \mu m^2$
Bi2212 mesa with $N \sim 200$ at $T_0 = 4.2 K$. a) IVC's of $3.5
\times 7.5 \mu m^2$ mesa with $N=10$ at $T_0=$ 4.9K, 40K, 60 K,
82.3K and 121 K. b) Temperature dependence of the superconducting
gap voltage.}
\end{minipage}
\end{figure}

\noindent branches due to sequential switching of IJJ's into the
resistive state  are clearly visible up to the largest current,
indicating that the mesa is not overheated above $T_c$ even at
such high $q$.

In inset a) of Fig. 2, IVC's at different $T_0=$ 4.9K, 40K, 60 K,
82.3K and 121 K, are shown for another mesa with small $N=10$ and
about an order of magnitude lower $q$. Estimations from Eq.(7)
yield that the self-heating in this case is 4-5 times less than
for the case shown in Fig. 2, i.e., $\Delta T$ should not exceed
20K at $T_0=4.9K$ \cite{Note}. Indeed, the IVC's are similar to
those obtained by short pulse measurements\cite{SuzukiIEEE} with
no sign of back-bending. The knee in the IVC's represents the
sum-gap voltage, $V_g=2N \Delta /e$, where $\Delta$ is the
superconducting energy gap. The temperature dependence of $V_g$
per junction for the case of small self-heating is shown in inset
b) of Fig. 2. We may use this $V_g (T)$ dependence for calibration
of self-heating at large $q$. For the case of Fig. 2, the $V_g$
per junction reduces to $\sim 35 mV$ at large bias, compared to
$\sim 52 mV$ in inset b). From this we estimate $\Delta T \sim 60
K$ at $V_g \sim 7 V $ in Fig. 2. Consequently $\Delta T$ at $V_g$
is $\sim 15K$ for the IVC at $T_0=4.9K$ in inset a) of Fig.2.

Our model uses an assumption that the heat escape from the mesa
occurs via diffusion into the bulk of the single crystal. This
assumption may become invalid when the mesa height $h$ is
$100-200$~\AA \ only, because the mean free path of low-frequency
phonons, which make an essential contribution to the heat
transport at low temperatures, can be $l_p\sim 1\ {\rm \mu m}\gg
h$~\cite{thermal_cond}. Such (non-equilibrium) phonons created in
one of the IJJ's would shoot through the whole thickness of the
mesa before releasing their energy via the electron-phonon and
phonon-phonon interactions already being deep in the bulk of the
single crystal. In other words there is an additional ballistic
heat channel, which would reduce the effective heat-power density
in the mesa. Thus our heat-diffusion estimations should be rather
considered as an $upper$ limit of the overheating.

In conclusion, we have shown analytically and numerically that
self-heating decreases in small mesa structures. Therefore,
reducing the mesa size is an effective way of avoiding
self-heating. We have demonstrated that self-heating can be made
tolerable in small HTSC mesa-structures even at high dissipation
power. This is important for future applications of HTSC mesas in
cryoelectronics and fundamental studies of HTSC. In general, the
obtained analytic solution is useful for better understanding of
thermal properties of small microelectronic elements.

\end{multicols}

\begin{references}
\bibitem{Note} Estimation from Eq.(6) yields somewhat larger value,
$\Delta T_m \simeq 33$ K for $q= 5 \times 10^3 W/cm^2$, $\langle a
\rangle = 2.5 \mu m$ and $\kappa_{ab}=\kappa_{ab}(T=T_m) \simeq 3
\times 10^{-2} Wcm^{-1}K^{-1}$. This shows that the heat
conductance via the electrode is significant.
\end{references}
\end{document}